# Role of Metal Centers in Tuning the Electronic Properties of Graphene-Based Conductive Interfaces

Silvio Osella,[1*†] Małgorzata Kiliszek,[2†] Ersan Harputlu,[3] Cumhur G. Unlu,[4] Kasim Ocakoglu,[5] Bartosz Trzaskowski,[1] Joanna Kargul[2*]

1. Chemical and Biological Systems Simulation Lab, Center of New Technologies, University of Warsaw, Banacha 2C, 02-097 Warsaw, Poland.

2. Solar Fuels Lab, Center of New Technologies, University of Warsaw, Banacha 2C, 02-097 Warsaw, Poland.

3. Advanced Technology Research & Application Center, Mersin University, Ciftlikkoy Campus, TR33343, Yenisehir, Mersin, Turkey.

4. Department of Biomedical Engineering, Pamukkale University, TR-20070 Denizli, Turkey.

5. Department of Energy Systems Engineering, Faculty of Technology, Tarsus University, 33400 Tarsus, Turkey.

CORRESPONDING AUTHOR: s.osella@cent.uw.edu.pl; j.kargul@cent.uw.edu.pl

† These authors contributed equally to the work.

**Abstract**

A major bottleneck in the fabrication of efficient bio-organic nanoelectronic devices resides in the strong charge recombination that is present at the different interfaces forming the complex system. An efficient way to overcome this bottleneck is to add a self-assembled monolayer (SAM) of molecules between the biological material and the electrode that promotes an efficient direct electron transfer whilst minimising wasteful processes of charge recombination. In this work, the presence of a pyrene-nitrilotriacetic acid layer carrying different metal centers as SAM physisorbed on graphene is fully described by mean of electrochemical analysis, field emission scanning electron microscopy, photoelectrochemical characterisation and theoretical calculations. Our multidisciplinary study reveals that the metal center holds the key role for the efficient electron transfer at the interface. While $Ni^{2+}$ is responsible for an electron transfer from SAM to graphene, $Co^{2+}$ and $Cu^{2+}$ force an opposite transfer, from graphene to SAM. Moreover, since $Cu^{2+}$ inhibits the electron transfer due to a strong charge recombination, $Co^{2+}$ seems the transition metal of choice for the efficient electron transfer.

**Introduction**

The creation of optically responsive materials is nowadays of high importance, since it allows to obtain a clean source of energy taking advantage of the conversion of light into electrons. Among the vast variety of methods developed over the years, the interfacing of light harvesting proteins (LHP) on a metal substrate is becoming more and more effective, thanks to technological advances in nanostructing the photoactive modules on the electrode surface.[1,2] In this devices, light is absorbed by the LHP, converted into electrons and transported to the electrode through the metal, in a variety of biotechnological applications, such as biosensors, biofuel cells, solar-to-fuel devices and biomolecular nanoelectronics.[3,4,5] Yet, despite the recent success in manufacturing such devices, a poor direct electron transfer (DET) is a major bottleneck, making the efficiency of such devices not competitive for practical applications.[6,7,8] The main factor responsible for a low DET is the charge recombination at the biological-metal interface, which strongly affect the overall performance of the devices. A second factor which impedes an efficient DET is the conformational flexibility of the LHP, which is essential in order to retain the protein function and ability. Thus, a deeper understanding of the transfer of charges at these critical interfaces in essential to overcome the DET efficiency limitations as well as the use of more complex interfaces in which at least three or more components are nanoengineered. To this end, a self-assembled monolayer (SAM) of well-ordered molecules is added to the LHP/metal interface to enhance the DET while maintaining the much-needed conformational flexibility to the protein.[9] Finally, the choice of the redox active metal center, that contributes to the final charge flow direction and efficiency, cannot be underestimated. Traditional metals used as electrode materials are gold or hematite,[10,11,12,13] but recently a new class of organic materials has been introduced as the next generation semimetal for bio-organic applications.

This semimetal of choice is graphene, a monolayer of $sp^2$ hybridized carbon atoms linked together to form a fully conjugated honeycomb lattice.[14,15] Its unique electronic properties, such as the linear dispersion of the valence and conduction bands at the high K symmetry point and the point degeneration of the two bands, make graphene an ambipolar material, able to efficiently transfer either electrons or vacancies with measured mobilities for a suspended monolayer exceeding 50000 $cm^2V^{-1}s^{-1}$ in ambient conditions.[16] The tunability of charge carriers and an impressive mobility make graphene the material of choice to enhance the DET in bio-organic interfaces. Yet, being a metal, graphene lacks a bandgap, which is needed for operating the electronic device. Many strategies have been developed over the years in order to open a bandgap in graphene, and among them, the addition of functional groups by either covalent or non-covalent interactions is one of the most robust and reliable approaches.[17,18,19,20,21]

One successful strategy to overcome the zero-bandgap problem is to build an interface in which the SAM is physisorbed on graphene, allowing for the opening of a gap in graphene and the increase of DET either from or to the graphene. Such a strategy provides fine tuning of the energy level of the frontier orbitals since the SAM-graphene interaction governs the charge flow direction. Moreover, if the SAM carries multiple functionalization groups, this 'orbital' effect can be strongly enhanced and finely tuned. The physisorption of SAM on the graphene surface allows the preservation of the high mobility of graphene by a non-covalent functionalization with molecules containing common π-systems, such as pyrene and its derivatives, to finally create a molecular interface between the functional molecules and the graphene surface.[17,18] As a consequence, on one hand the π-conjugated structure of graphene is preserved and a bandgap is opened but, on the other hand, the reversibility of the interaction can result in the desorption of the molecules from the graphene surface. Pyrene-derivative molecules have been extensively used as good candidates for organic conductive interfaces containing redox active catalysts and photoactive enzymes, after their immobilization on graphene.[22]

In this paper, we designed and studied, both theoretically and electrochemically, several types of single layer graphene (SLG) devices containing pyrene derivatives functionalized with a nitrilotriacetic acid group (NTA), coordinated with different divalent metal cations ($M^{2+}$): $Ni^{2+}$, $Co^{2+}$ and $Cu^{2+}$ (SLG/pyrNTA-$M^{2+}$, see **Figure 1**). The aim of our study was to determine the role of different metal centers present within the pyrene SAM for an improved DET. Nowadays the most common metal center used for oriented binding of His$_6$-tag engineered proteins is Ni$^{2+}$. Nevertheless, considering an improved efficiency of DET within nanostructured bioelectronic devices, other metal centers, such as the neighbouring metals in the periodic table, should be explored. Yet, both Co$^{2+}$ and Cu$^{2+}$ cations present a radical character which might, in turn, strongly change the final DET output at the interface. Therefore, we first applied density functional theory (DFT) calculations to assess the change in the work function and the direction of the charge transfer (CT) flow depending on the different metal cation used. We then verified our theoretical CT models with electrochemical investigations of the constructed SLG/pyrNTA-M$^{2+}$ electrodes. We would like to emphasize that the final device obtained is a full-solid state device in which no solvent is present. Our DFT calculations showed that despite a similar positive shift of the graphene work function in the case of either Co or Cu redox centers, the DET from graphene to the SAM is enhanced only when the Co$^{2+}$ is considered, while a strong charge recombination is found for the Cu$^{2+}$-containing interface. An opposite CT direction, i.e. from SAM to graphene, is observed when the Ni$^{2+}$ cation is coordinating the pyrNTA SAM.[23] The electrochemical data confirm the theoretical CT models and stress the importance of the imidazole molecule, used to complete the coordination sphere of the M$^{2+}$ center, as the main attenuator of charge flow.

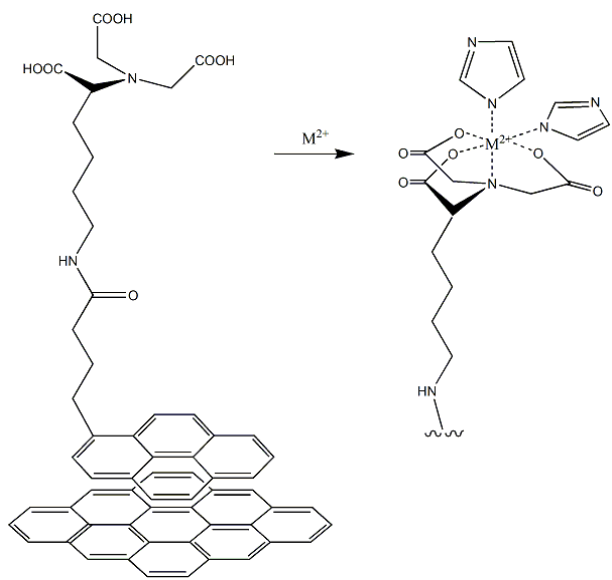

**Figure 1.** Chemical structure of the graphene/SAM interface studied. $M^{2+} = Ni^{2+}$, $Co^{2+}$ and $Cu^{2+}$.

**Methodology**

Quantum mechanical calculations

In the present study we considered a pyrene-NTA-$M^{2+}$ system with three different metal centers: $Co^{2+}$ (pyrNTA-Co-IM), $Ni^{2+}$ (pyrNTA-Ni-IM) and $Cu^{2+}$ (pyrNTA-Cu-IM), physisorbed as SAM on SLG. A similar set up as reported in our previous study has been adopted.[23] The geometry optimization of the full, periodic interfaces has been performed at the DFT level of theory, with the PWscf package of the Quantum Espresso suite of programs.[24] PBE functional coupled with the vdw-DF2 term[25] for the exchange and correlation was used to account for van der Waals interactions together with ultrasoft pseudopotentials[26] with a cut-off of 50 and 200 Ry for the expansion of the wave function and density, respectively. Since PBC conditions are used, the final structure should be neutral; thus, to assure a null net charge, one carboxyl group has been protonated. As a consequence, the geometry of the system is now distorted and does not reflect the expected octahedral coordination, but a square planar geometry. Although this might not be the lowest energy conformation of the SAM, all three interfaces have been built in the same way, thus assuring consistency for the calculations. To consider the radical nature of the systems when Co and Cu are present, spin magnetization was included during the calculations. After optimization at the DFT level of theory with the PBE functional, the distance between pyrene and graphene was measured to be between 3.7 and 3.8 Å. The work function analysis has been carried considering the methodology reported in our previous study.[23] Briefly, the averaged electrostatic potential along the axis normal to the interface (i.e. z-axis) is used to directly estimate the work function shift, comparing the potential on the bare side and on the SAM-covered side of the surface. The total work function is casted down into two contributions:

$$\Delta\Phi = \Delta\Phi_{SAM} + \Delta\Phi_{CT} \qquad (1)$$

Where the molecular contribution ($\Delta\Phi_{SAM}$) arises from the dipole moment of the SAM backbone, and the charge transfer contribution ($\Delta\Phi_{CT}$) from the interfacial electronic reorganization upon physisorption of the SAM on graphene. $\Delta\Phi$SAM is obtained by computing the electrostatic potential profile across the molecules without graphene, while keeping the coordinates of the system frozen. The $\Delta\Phi$CT contribution is then the difference of the two terms. Only when a dipole moment is present in the SAM the shift in the work function is observed.

FTO/SLG electrode functionalization with the pyrene-NTA-$M^{2+}$ SAM

The SLG/FTO electrodes were firstly modified with pyrene-nitrilotriacetic acid moiety according to the protocol described in detail in Osella et al.[23] Various metal cations were coordinated to the pyrNTA moiety during 1-hour incubation at RT in 100 mM $Co(NO_3)_2$, $NiSO_4$ and $CuSO_4$ solutions. For imidazole modification, the SLG/PyrNTA-M-IM electrodes were incubated in 100 mM $C_3H_4N_2$ solution for 45 min at RT. Elemental analysis of the electrodes was performed by a field emission scanning electron microscopy (FE-SEM, Zeiss/Supra 55) connected with energy dispersive X-ray spectrometer (EDX), as described in [23]. Existence of the pyrNTA-Co, pyrNTA-Ni and pyrNTA-Cu molecules on the SLG/FTO surface was confirmed by X-ray photoelectron spectroscopy (XPS) measurements, performed on a Specs-Flex mode instrument.

Electrochemical measurements

A Versa STAT 3 (Princeton Applied Research, USA) electrochemical workstation equipped with a KL 2500 LCD halogen white light source (Schott, Germany) was used to perform photoelectrochemical analysis. A CHI 6502D potentiostat was utilized for cyclic voltammetry (CV) and differential pulse voltammetry (DPV) measurements. A custom-built Teflon three-electrode cell filled with 5 mM Ar-saturated phosphate buffer (pH 7) was utilized for the photoelectrochemical characterisation. An Ag/AgCl (3 M KCl) electrode and glassy carbon rod were used as the reference (REF) and counter (CE) electrodes, respectively. For photochronoamperometric analysis of SLG/pyrNTA-M and SLG/pyrNTA-M-IM samples the FTO served as working electrode (WE), whereas for the CV and DPV analysis a glassy carbon electrode, respectively. The open circuit potential (OCP) was recorded in the dark conditions. Photochronoamperometric experiments were performed at different potentials for each electrode under the light intensity of 100 mW·cm$^{-2}$, with light ON/OFF periods of 30 seconds. The surface coverage ($\Gamma$) of metal centers can be estimated from the charge using the following equation:

$$\Gamma = Q / nFA \qquad (2)$$

where $Q$ is the electric charge obtained by integration of area under the reduction peak of CVs curves, $n$ is the number of electrons transferred in the redox reaction (n=1) and $F$ is the Faraday constant (9.648 × 10$^4$ C·mol$^{-1}$), and A is the geometric area of the electrode (0.4185 cm$^2$).

**Results and discussion**

**Metal coordination alters charge transfer properties of the SLG/pyrNTA-M-IM assemblies**

After the geometry optimization, the whole SAM structure assumes an elongated conformation on the graphene surface, with an all 'anti' conformation of the alkyl backbone, and with pyrene oriented on the graphene surface in a AA-stacking pattern at an optimized distance of 3.7 Å – 3.8 Å (see Supporting Information for more details). The adsorption energy has been obtained as:

$$E_{ads} = E_{SLG/SAM} - (E_{SLG} + E_{SAM}) \qquad (3)$$

where the E$_{SLG}$ and E$_{SAM}$ are the contributions of the two components of the system calculated at the optimized geometry of the entire system. Adsorption energy values of -0.90 eV, -0.45 eV and -1.13 eV were found for SLG/pyrNTA-Co-IM, SLG/pyrNTA-Ni-IM and SLG/pyrNTA-Cu-IM, respectively, ensuring the stability of the complexes on the graphene surface.

The plane average potential profiles of the three interfaces are depicted in **Figure 2**. The calculated work function for the whole interface is equal to 5.92 eV, 5.01 eV and 5.74 eV for SLG/pyrNTA-Co-IM, SLG/pyrNTA-Ni-IM and SLG/pyrNTA-Cu-IM interfaces, respectively. These show an increase of 1.34 eV, 0.44 eV and 1.16 eV with respect to bare graphene surface work function, with a computed value of 4.58 eV, in good agreement with experimental measurements of 4.6 eV.[27]

To gain a deeper insight into the different contribution governing the work function shift, the molecular backbone ($\Delta\Phi_{SAM}$) and CT ($\Delta\Phi_{CT}$) components were separately investigated, to quantify the contribution from the molecular dipole moment *versus* the supramolecular effect arising from the transfer of charges at the interface. The shifts arising from the free SAM, shown in **Figure 2**, have values of 0.85 eV, 1.05 eV and 0.70 eV for pyrNTA-Co-IM, pyrNTA-Ni-IM and pyrNTA-Cu-IM, respectively. The supramolecular effect arising from the charge transfer contribution was found to be equal to 0.36 eV, -0.52 eV and 0.31 eV for SLG/pyrNTA-Co-IM, SLG/pyrNTA-Ni-IM and SLG/pyrNTA-Cu-IM interfaces, respectively. Here, two different effects can be distinguished. First, the $\Delta\Phi_{SAM}$ contribution for the SLG/pyrNTA-Ni-IM interface is much stronger compared to the assemblies with $Co^{2+}$ or $Cu^{2+}$ cations, despite the smaller dipole moment calculated (see below). Second, while for $Co^{2+}$ and $Cu^{2+}$ the $\Delta\Phi_{SAM}$ and the $\Delta\Phi_{CT}$ contribution act to enhance the total work function shift, the presence of $Ni^{2+}$ alters the trend with an opposite interaction, which is the main factor responsible of the smaller total calculated work function shift.

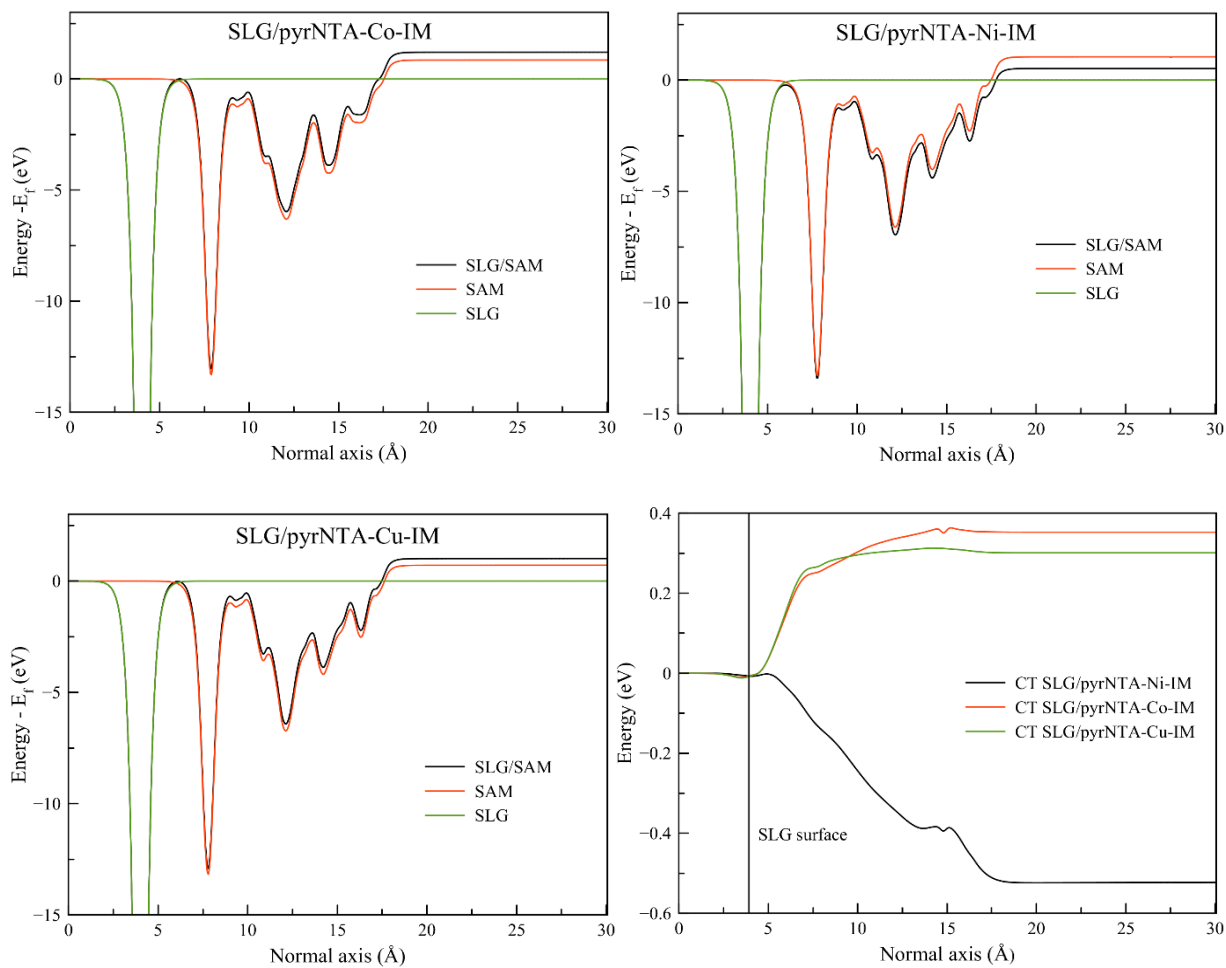

**Figure 2.** Plane averaged potential of the SLG/pyrNTA-Co-IM, SLG/pyrNTA-Ni-IM and SLG/pyrNTA-Cu-IM interfaces. The bare graphene monolayer potential is shown in green, the graphene surface covered by SAM in black and the free SAM layer in red. The evolution of the charge transfer contribution to the work function is also reported (bottom, right). From left to right in each plot, we move from the graphene surface to the SAM contribution, away from the surface.

The dipole moment of the SAM is responsible for the molecular backbone contribution. Surprisingly, for the SLG/pyrNTA-Ni-IM the value of dipole moment component perpendicular to the graphene was found to be -1.83 D, while the presence of different $M^{2+}$ cation strongly enhanced it to -4.23 D and -3.55 D for SLG/pyrNTA-Co-IM and SLG/pyrNTA-Cu-IM interfaces, respectively. The strong difference (up to 2.4 D) in dipole moment values can be attributed to the different nature of the $M^{2+}$ cation, which is the key to quantify the contribution of $\Delta\Phi_{SAM}$ to the total work function. In fact, although all three metal centers considered in the study are the first row transition metals and neighbours in the periodic table, $Co^{2+}$ and $Cu^{2+}$ are formally radicals, which may contribute to the strong difference in dipole moment calculated for each SAM structure used in this study. **Table 1** summarizes the results of the work function analysis.

**Table 1.** Work function analysis casted down among the three components, $\Delta\Phi$ $\Delta\Phi_{SAM}$ and $\Delta\Phi_{CT}$. The dipole moment of the SAM calculated along the z-axis is also reported.

| Interface | $\Delta\Phi$ (eV) | $\Delta\Phi_{SAM}$ (eV) | $\Delta\Phi_{CT}$ (eV) | Dipole moment (D) |
|---|---|---|---|---|
| SLG/pyrNTA-Co-IM | 1.21 | 0.85 | 0.36 | -4.23 |
| SLG/pyrNTA-Ni-IM | 0.53 | 1.05 | -0.52 | 1.83 |
| SLG/pyrNTA-Cu-IM | 1.01 | 0.70 | 0.31 | -3.55 |

The second important parameter to consider is the CT contribution $\Delta\Phi_{CT}$, which is substantial, and which has different sign for the different metal centers. As observed in our previous work,[23] the CT is not confined to the graphene/SAM interface ligated with various $M^{2+}$ cations (*i.e.* interaction between graphene and pyrene) but increases until the end of the molecular backbone (**Figure 2**). The plot illustrates that the CT contribution is similar for the two systems when the $Co^{2+}$ and $Cu^{2+}$ metal cations are considered, while it is negative when the $Ni^{2+}$ cation is present within the molecular backbone. This effect is translated into a different direction of charge flow and can be quantified by considering the unbalance of charge between the two components of the interface in their ground state. The CT contribution translates into a (partial) transfer of charges form one fragment to the other, depending on the metal center. Bader charge analysis allows to calculated the excess/depletion of charge at the interface as:

$$\Delta\rho(z) = \rho_{SLG/SAM}(z) - [\rho_{SLG}(z) + \rho_{SAM}(z)] \qquad (4)$$

were $\rho_{SLG/SAM}$ is the charge density of the full system and $\rho_{SLG}$ and $\rho_{SAM}$ the charge densities on the two non-interacting fragments. We found an excess of electrons of -0.08 |e| on the SAM for both SLG/pyrNTA-Co-IM and SLG/pyrNTA-Cu-IM interfaces, while for the system with the $Ni^{2+}$ cation this value is reversed, with an excess of electrons (-0.05 |e|) on the graphene surface. Thus, when the coordination of $Ni^{2+}$ is considered, the charge flow is opposite, going from the SAM to graphene. Thus, we suggest that the strong positive shift of the work function for both systems with $Co^{2+}$ and $Cu^{2+}$ coordination arises from the synergic interaction of the CT and the molecular backbone contributions, which act to enhance the overall work function shift, resulting in an electron flow from graphene to the SAM. On the other hand, for the SLG/pyrNTA-Ni-IM interface these contributions counteract due to the different nature of the metal cation, resulting in the opposite direction of the electron flow, i.e., from SAM to the graphene.

To gain an insight into the MLCT process, and to validate our hypothesis, the analysis of the density of states (DOS) was performed. Total DOS and the projection over the fragments (SAM and SLG) as well over atomic types are presented in **Figure 3.**

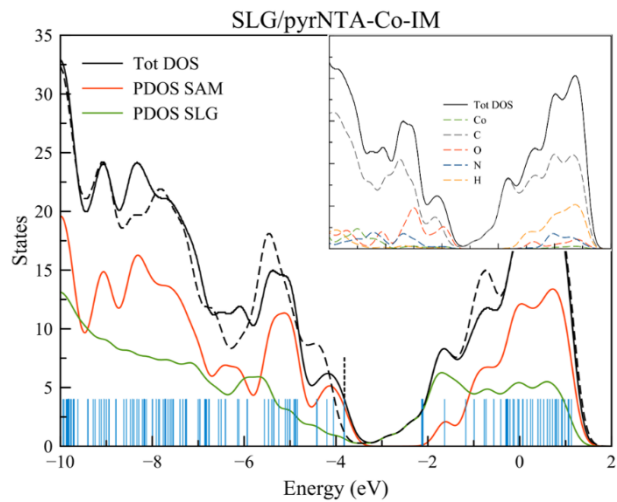

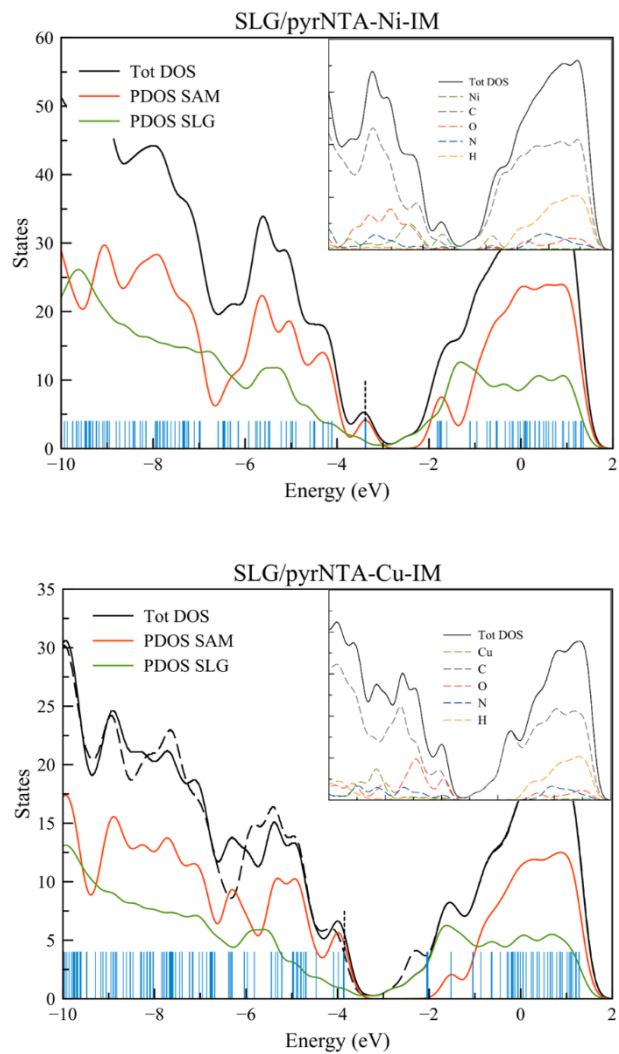

**Figure 3.** Total density of states and projected density of states over the fragments and over each atom (inserts) for the three interfaces analysed. Vertical blue lines indicate the eigenvalues; dotted line indicates the position of the Fermi energy level. Black dashed line refers to the total DOS calculated for the 'spin down' electrons for the two open shell systems.

As already observed for the previous analysis, two different scenarios arise depending on the different metal cations considered. For both systems with $Co^{2+}$ and $Cu^{2+}$ metal ions, the states pinned at the Fermi level (-4.06 eV and -3.99 eV for SLG/pyrNTA-Co-IM and SLG/pyrNTA-Cu-IM interfaces, respectively) are very close in energy levels to the valence band maximum level (VBM) arising from the presence of an open shell system, with energy at -3.82 eV and -3.87 eV for SLG/pyrNTA-Co-IM and SLG/pyrNTA-Cu-IM interfaces, respectively. Both levels are therefore contributing to the Fermi energy, while for the Conduction Band Minimum (CBM), lying at -2.12 eV and -2.04 eV for SLG/pyrNTA-Co-IM and SLG/pyrNTA-Cu-IM, respectively, there is a sole contribution from the graphene surface. Interestingly, the LUMO of the SAM is located at much higher energy with respect to the CBM, at -1.63 eV and -1.52 eV for pyrNTA-Co-IM and pyrNTA-Cu-IM, respectively. From the atomic PDOS, we observe that the peak at the Fermi level is a sum of different contributions from carbon and oxygen atoms, while the CBM is purely that of carbon DOS. Additionally, the shape of the crystal orbitals (depicted in **Figure S4**) also confirms the localization of the Fermi level and the VBM level over the NTA moiety of the SAM, with the CBM of the system is delocalized over the graphene layer. Interestingly, the LUMO of the SAM is localized over the pyrene moiety of the molecule. The strong different localization of this frontier levels is the main phenomenon responsible for the high CT observed for these interfaces.

A different scenario occurs when the SLG/pyrNTA-Ni-IM interface is considered (**Figure 3**). In this case the contribution at the Fermi energy level is due to the SAM only, while the CBM the contribution is mixed, arising from both SAM and graphene. In striking contrast with respect to the previous interfaces, the PDOS analysis reveals that the $Ni^{2+}$ cation is the main element responsible for the VBM and CBM peaks, together with a strong contribution of the carbon atoms. Again, the shape of the orbitals involved in these levels clarify the different behaviour of this interface. In fact, now the Fermi level (which correspond to the VBM), lying at -3.38 eV, is characterized by one orbital localized over the NTA moiety as well as over the $Ni^{2+}$ metal center, while for the conduction band the situation is completely different. A double contribution is now present, with the LUMO of the SAM lying at -1.81 eV and the CBM very close in energy, at -1.77 eV, but while the first is fully localized over the NTA and $Ni^{2+}$ parts of the SAM, the latter is delocalized over the graphene layer and the pyrene group of the SAM (**Figure S4**).

In conclusion, different directionalities of the charge flow observed for the three distinct SLG/pyrNTA-M-IM interfaces arise from the different nature of the metal center (radical species or closed shell system) which, in turn, determines the energy level alignment at the interfaces. The (P)DOS and MOs analysis (different localization of the orbitals over different parts of the interface) help rationalizing the relative order of the MOs once the interface is built, and the magnitude of supramolecular CT effect. From the Bader charge analysis, we have more insights into the flow of electrons. In particular, the Fermi of the SAM molecule is higher in energy compared to the Fermi of graphene for the SLG/pyrNTA-Co-IM and SLG/pyrNTA-Cu-IM interfaces, leading to a final excess of electrons over the SAM fragment. Conversely, for the SLG/pyrNTA-Ni-IM interface the Fermi level of graphene is higher in energy, thus an opposite flow of electrons, from SAM to graphene, is obtained.

**Elemental and photoelectrochemical characterisation of the metal-coordinated SLG/pyrNTA-M-IM assemblies**

To experimentally investigate the electron transfer properties of the three $Me^{2+}$-coordinated SAMs assembled on the SLG surface, we prepared the relevant electrodes on the FTO/SLG surface. Firstly, we performed the quantitative elemental analysis of the SAMs on the FTO/SLG surfaces using energy dispersive X-ray spectroscopy (EDX) to confirm the presence and assess the coverage of SLG with pyrNTA-Co-IM, pyrNTA-Ni-IM and pyrNTA-Cu-IM SAM molecules. Before EDX analysis, the samples were covered with a very thin layer of Pt by using a Pd/Pt alloy target. This step is necessary to image a nonconductive sample at higher voltages, and it can explain the presence of platinum and palladium observed in the EDX spectra. The EDX mapping of all the three assemblies confirmed at an atomic scale the presence of SAM molecules on all FTO/SLG surfaces, showing rather regular distribution of the Co, Ni, Cu and N atoms (0.46% Co, 0.58% Ni, 0.61% Cu and ≈ 0.96% N) on the SLG surface (see **Figure 4**). In addition, the other elements which were observed in EDX spectra (Sn, O, Si and Pt/Pd) mostly come from FTO glass and coating process of a conductive layer.

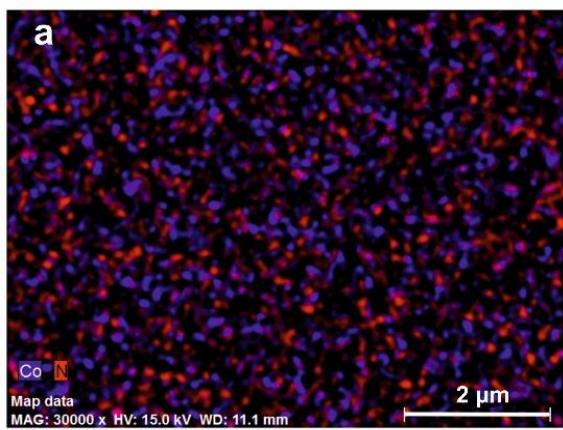 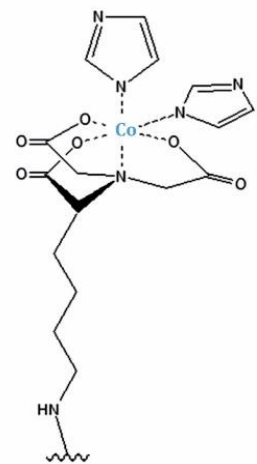

| Element | Weight % |
|---|---|
| Sn | 51.41 |
| Pt | 19.51 |
| Pd | 7.27 |
| O | 5.78 |
| C | 3.80 |
| Si | 2.55 |
| N | 0.92 |
| Co | 0.46 |

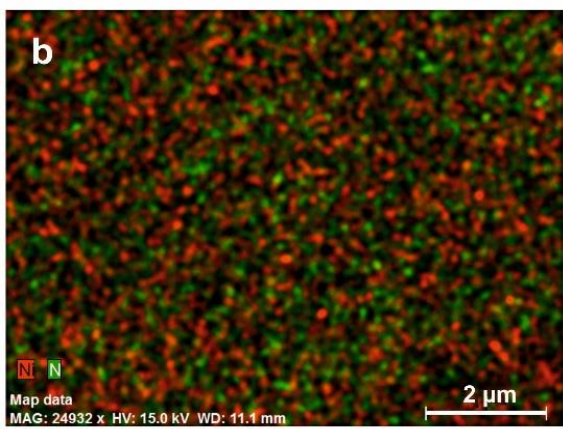 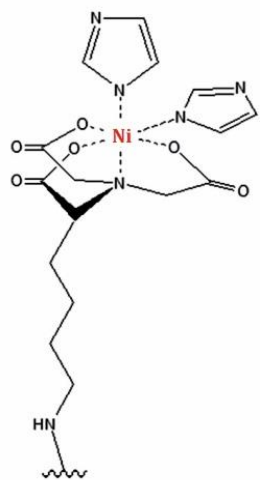

| Element | Weight % |
|---|---|
| Sn | 64.81 |
| Pt | 13.48 |
| Pd | 8.14 |
| O | 5.80 |
| C | 3.56 |
| Si | 2.57 |
| N | 0.99 |
| Ni | 0.58 |

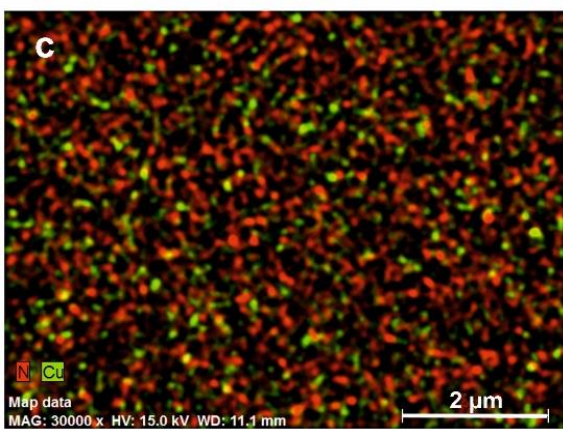 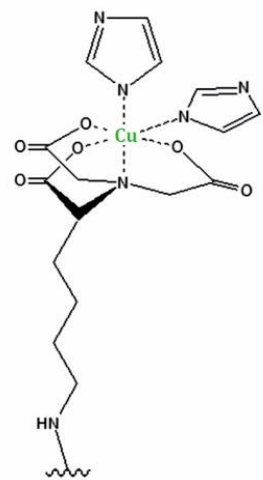

| Element | Weight % |
|---|---|
| Sn | 47.58 |
| Pt | 14.41 |
| Pd | 6.56 |
| O | 6.02 |
| C | 3.16 |
| Si | 2.26 |
| N | 0.98 |
| Cu | 0.61 |

**Figure 4.** EDX atomic mapping of the pyrNTA-Me$^{2+}$-IM modified SLG electrodes. Blue (a), red (b) and green (c) spots represent the Co$^{2+}$, Ni$^{2+}$ and Cu$^{2+}$ ions, respectively, present on the SLG electrode surface. The red (a,c) and green (b) spots represent the N atoms that are present within the moieties. *Right*: Quantitative elemental analysis of each selected EDX maps and chemical structures of all the molecules forming SAMs on SLG.

The broad-scan XPS analysis of pyrNTA-M (Co-Ni-Cu)-functionalized SLG on the FTO surface is shown in **Figure 5**.

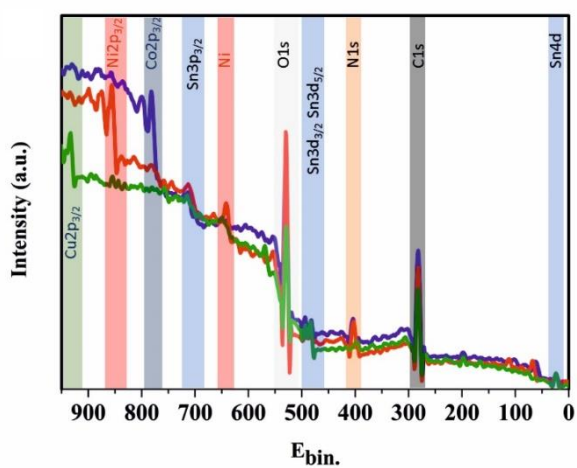

**Figure 5.** Broad-scan XPS spectra of FTO/SLG/pyrNTA/Ni-Co-Cu surfaces (blue line represented FTO/SLG/pyrNTA-Co, red line represented FTO/SLG/pyrNTA-Ni and green line represented FTO/SLG/pyrNTA-Cu).

The XPS spectra reveal the presence of carbon, nitrogen and cobalt in the SLG/pyrNTA-Co and the signals corresponding to C1s, N1s, Co2p3/2 were determined at 283.1 eV, 403.3 eV and 780 eV, respectively (**Figure S5** a1, a2, a3). Carbon, nitrogen and nickel in the SLG/pyrNTA-Ni and the signals corresponding to C1s, N1s, main Ni2p3/2 were determined at 283.4 eV, 402.9 eV and 853 eV, respectively (**Figure S5** b1, b2, b3). Finally, the presence of carbon, nitrogen and copper in the SLG/pyrNTA-Cu and the signals corresponding to C1s, N1s, main Cu2p3/2 were determined at 283.4 eV, 402.9 eV and 933.2 eV, respectively (**Figure S5** c1, c2, c3). In details, C1s (sp2) contribution (**Figure S5** a1, b1, c1) is the main component for SLG and pyrene moieties, N is a main element in the NTA molecule and the presence of N1s additive is clearly shown in **Figure S5** a2, b2, c2. The signals corresponding to Sn3d5/2 and O1s which originate mostly from FTO surface, were determined at around 484 eV and 531 eV, respectively.

In order to characterize electrochemical properties of the designed systems, cyclic voltammetry (CV) in conjunction with differential pulse voltammetry (DPV) approach was used (see **Figure 6**). Average Γ values (see Eq. 3) were estimated as $1.26 \cdot 10^{-11}$, $1.18 \cdot 10^{-11}$ and $3.51 \cdot 10^{-12}$ mol·cm$^{-2}$ for Cu, Ni and Co metal centers, respectively, indicating ultrathin layer adsorption for each metal ion-pyrene film. Similar Γ values were reported for various (bio)organic and inorganic monolayers on various electrode materials including graphene.[28,29,30,31]

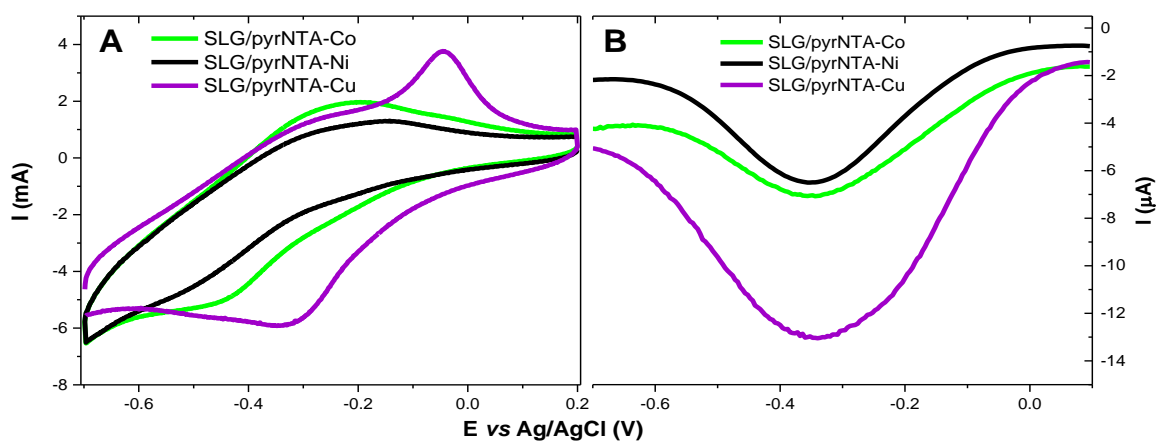

**Figure 6.** Electrochemical characterization of the pyrNTA-Me$^{2+}$ modified SLG electrodes. (A) Cyclic voltammograms for the SLG/pyrNTA-M systems recorded at 50 mV/s in Ar-saturated 5 mM phosphate buffer (pH 7.0); (B) differential pulse voltammograms for the respective electrodes recorded in Ar-saturated 5 mM phosphate buffer (pH 7.0); DPV parameters: pulse amplitude 50 mV, pulse width 50 ms.

The CV characterization clearly shows a well-defined pair of redox peaks for the SLG/pyrNTA-Cu electrode. These two redox peaks, appearing at around -0.05 V in the anodic and -0.3 V at the cathodic scan, correspond to reversible $Cu^{II}/Cu^{I}$ redox couple, as shown previously.[32] The voltammetric profiles for the SLG/pyrNTA-Co and the SLG/pyrNTA-Ni systems also illustrate peaks characteristic for oxidation and reduction of particular active metal centres, albeit their definition is somewhat less defined compared to the Cu-based system. Nevertheless, the electrochemical signals for redox $Co^{II}/Co^{III}$ couple appear at around -0.25V and -0.45V, respectively, as well as for $Ni^{II}/Ni^{III}$ couple, at around -0.2 and -0.5V, respectively. In order to obtain a deeper insight into electrochemical properties of the redox pairs investigated in this study, a DPV analysis was carried out for each system (**Figure 6B**). This type voltammetry is much more sensitive than CV and it is used in the present study to confirm the presence of a very low surface concentration of metal centers integrated with pyrene-NTA moiety. Indeed, differential pulse voltammograms showed well-defined redox active peaks obtained for the three analyzed electrodes that are derived from the respective metal centres coordinated to pyrene-NTA groups. Figure S6 presents the photocurrents recorded from all the three electrode configurations containing $Cu^{2+}$, $Co^{2+}$ or $Ni^{2+}$ redox centers during prolonged illumination. The data confirms that the photocurrents are rather stable up to at least 1 hour of continuous illumination. The SLG/pyrNTA-Cu electrode is generated the highest currents whilst similar smaller current values were obtained from the SLG/pyrNTA-Co and SLG/pyrNTA-Ni nanosystems. This data also confirms the CV and DPV data recorded for the three electrode nanoarchitectures (see **Figure 6**).

In order to experimentally verify our DFT modelling of the directionality of electron flow within the SLG/pyrNTA-M assemblies, photochronoamperometric measurements were conducted on all types of the functionalized SLG electrodes. **Figure 7** presents the comparison of the current densities (j) at various potentials, obtained from the graphene monolayer modified with pyrNTA SAM coordinated with three distinct $M^{2+}$ cations, in the absence of (SLG/pyrNTA-M) (**Figure 7A**) and presence of imidazole (SLG/pyrNTA-M-IM) (**Figure 7B**).

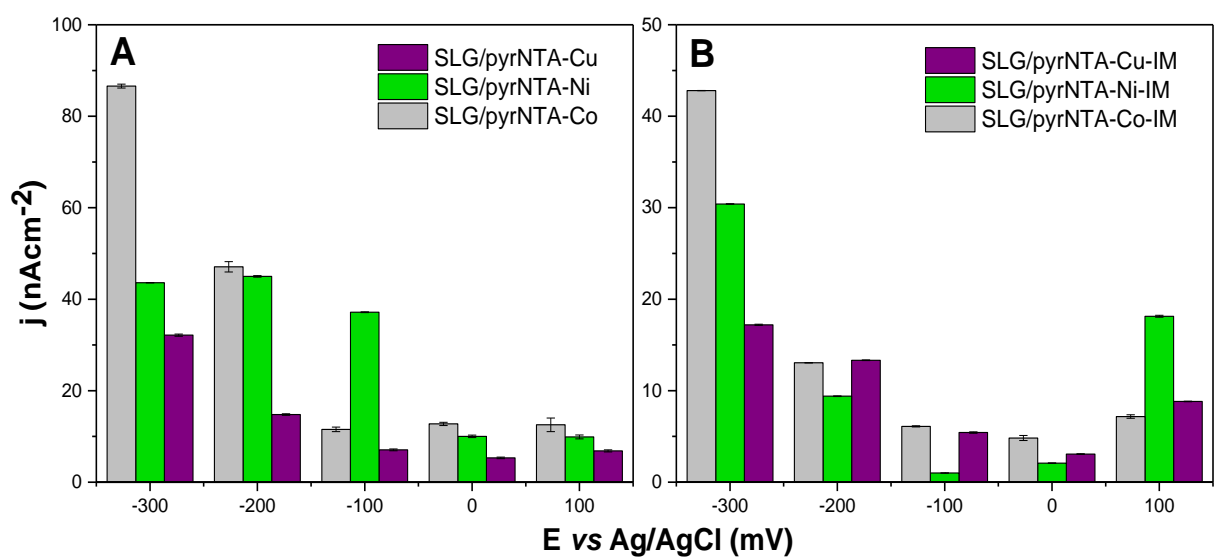

**Figure 7.** Photochronoamperometric analysis of SLG/pyrNTA-M and SLG/pyrNTA-M-IM assemblies. Photocurrent densities generation at respective electrodes: (A) without imidazole (SLG/pyrNTA-M) and (B) with imidazole (SLG/pyrNTA-M-IM) as a function of external potential. The average of the current densities values was estimated from two independent measurements (n = 2).

The electrochemical data confirms the results of our DFT calculations of the work function shift affecting the directionality of the charge flow depending on the coordinated metal center. Overall, the cathodic photocurrent generation from the three types of electrode assemblies decreases in the following order: SLG/pyrNTA-Co > SLG/pyrNTA-Ni > SLG/pyrNTA-Cu (**Figure 7**). At an open circuit potential (OCP), the highest current of 12.8 nA·cm$^{-2}$ was obtained from the SLG/pyrNTA-Co electrode compared to SLG/pyrNTA-Ni (10 nA·cm$^{-2}$) and SLG/pyrNTA-Cu (5.3 nA·cm$^{-2}$) samples in the absence of imidazole (**Figure 7A**). On the other hand, when the coordination sphere of the metal centers is completed with imidazole the trend is reversed for Ni and Cu ions, and now reads pyrNTA-Co-IM > SLG/pyrNTA-Cu-IM > SLG/pyrNTA-Ni-IM (**Figure 7B**). The photocurrents were clearly enhanced when a more negative bias was applied. At -300 mV the current density value was 86.6 nA·cm$^{-2}$ for the SLG/pyrNTA-Co electrode, which represents over 2.5-fold increase of the cathodic photocurrents compared to the SLG/pyrNTA systems coordinated with Ni$^{2+}$ and Cu$^{2+}$ cations, with values of 43.6 and 32.1 nA·cm$^{-2}$ respectively.

The same mode of electron transfer, from graphene to the pyrNTA SAM, should be observed for the SLG/pyrNTA-Cu interface. Indeed, at an OCP as well as negative bias of -100 mV the currents generated from the Co- and Cu-containing samples were similar (see **Figure 7A**). However, the currents obtained for the Cu-containing assemblies were generally lower compared to the Co-samples, indicating the presence of charge recombination processes occurring within SAM-Cu, especially at a high negative bias (see **Figure 7A**).

Similar characteristics of photocurrent generation can be observed when the coordination sphere of the metal center is completed with imidazole. Overall, imidazole modification of SLG/pyrNTA films resulted in considerable lowering of the photocurrents for all the electrode assemblies (**Figure 7B**), in agreement with our previous study.[23] At an OCP, the highest currents were observed for the samples with Co$^{2+}$ (4.8 nA·cm$^{-2}$), the lowest ones for the assemblies with Ni$^{2+}$ (2.1 nA·cm$^{-2}$) and an intermediate situation when Cu$^{2+}$ is present (3.1 nA·cm$^{-2}$). Similarly, the highest cathodic photocurrents were recorded for the samples with Co$^{2+}$ (43 nA·cm$^{-2}$ at -300 mV), and the lowest ones for Cu-containing samples (17.2 nA·cm$^{-2}$, see **Figure 7B**). On the other hand, the highest anodic photocurrent densities were recorded for the samples with Ni$^{2+}$ and imidazole (18.1 nA·cm$^{-2}$), confirming that Ni$^{2+}$ cation promotes the charge transfer from SAM to graphene, as already observed in [23]. Nevertheless, the redox behaviour of the assemblies is more complex in the presence of $Ni^{2+}$ cation compared to the other metal centers investigated in this study, as this cation seems to promote generation of also cathodic currents at a negative bias of at least -200 mV (see **Figure 7B**). Moreover, a similar behaviour is observed for the Cu- and Co-containing interfaces, in which an enhancement of the anodic current is measured, up to 8.8 and 7.2 nA·cm$^{-2}$, which is in apparent contrast with the charge flow direction. Both behaviors can be explained by the unbalance between the DET and the overpotential applied, in which the last one overturns the (small) ground state charge transfer for the cathodic current and decreases the electron injection barrier by applying negative external field. For generation of the anodic currents, the situation is more complex, since the unexpected trend is observed only for Co- and Cu-containing interfaces; yet, the explanation holds true also in this case and smaller external electric field might be applied.

**Conclusions**

In this study we report the quantum mechanical and electrochemical characterizations of three various SLG assemblies functionalized with pyrNTA moiety conjugated with three different metal centers ($M^{2+}$: $Co^{2+}$, $Ni^{2+}$, $Cu^{2+}$) exerting different effects on the Fermi energy levels of the conductive pyrNTA SAM. Both theoretical and electrochemical data confirm the strict relationship between the directionality of the charge flow between graphene and pyrNTA interface and the electronic properties of the coordinated metal center. In particular, we observed that the SLG/pyrNTA-Ni-IM interface favours the electron transfer from SAM to graphene, while the presence of the other two metal cations favours an opposite flux of electron, from graphene to SAM. In addition, the presence of $Cu^{2+}$ hampers the DET due to strong charge recombination at the interface, while $Co^{2+}$ seems an ideal metal of choice.

Our combined quantum and electrochemical data points towards the rational design for the optimized functionalized graphene photoelectrode, whereby careful selection of the metal center has the profound effect on the preferred directionality of the electron flow. This study clearly shows that the most promising system for the pyrene-NTA-SLG photocathode assembly should incorporate $Co^{2+}$ cation as the ligand to NTA for the most efficient graphene-to-SAM charge transfer. On the other hand, for the opposite CT (from pyrNTA SAM to graphene) $Ni^{2+}$ cation should be utilized, since it promotes the highest anodic photocurrents in accordance with the electrochemical data and DFT calculations of the energy levels.

In summary, the present work paves the way for the optimal design of the highly oriented biophotoelectrode assemblies (incorporating $His_6$-tag as the protein binding site), in which the directionality of the electron flow can be fine-tuned within the conductive interface composed of the pyrNTA moiety by introducing the specific metal redox center depending on the desired configuration of the electrode.

**Supporting Information**

Geometrical analysis of the three interfaces, Cartesian coordinates of the optimized systems at the DFT level of theory, shape of the frontier molecular orbitals and XPS spectra. The Supporting Information is available free of charge on the ACS Publications website.

**Acknowledgements**

SO acknowledges the financial support from the Polish National Science Center, grant UMO-2015/19/P/ST4/03636 (POLONEZ 1) for the funding from the European Union's Horizon 2020 research and innovation program under the Marie Skłodowska-Curie grant agreement No. 665778. JK and MK acknowledge the financial support from the Polish National Center for Research and Development (grant no. DZP/POLTUR-1/50/2016, agreement no. 5/POLTUR-1/2016 to JK), while KO and EH are supported by the Scientific and Technological Research Council of Turkey,

TÜBITAK (grant no. 215M389 to KO) within the framework of 1st Bilateral Polish-Turkish POLTUR program. We thank Sebastian Maćkowski (Nicolaus Copernicus University, Poland) and his team for helpful discussions and collaboration. We are also grateful to Renata Bilewicz and Agnieszka Więckowska (Faculty of Chemistry, University of Warsaw) for many helpful discussions and access to some of their electrochemical equipment important for this study. The calculations were partially performed at the Interdisciplinary Center for Mathematical and Computational Modelling (ICM, University of Warsaw) under the G53-8, GA-69-26 and GA73-16 computational grants.

**TOC Graphic**

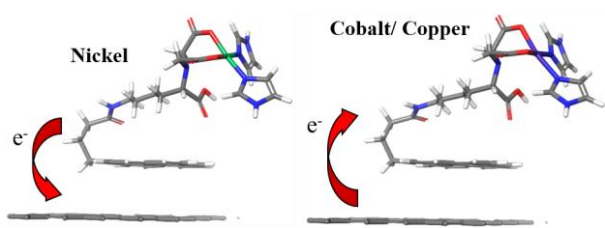